\documentclass{PoS}

\title{Transient Emissions from Radio-Active Stars:  Implications for Wide-field Radio Surveys}

\ShortTitle{Radio-Active Stars}

\author{\speaker{Rachel A. Osten}\thanks{Hubble Fellow}\\
        University of Maryland\\
        E-mail: \email{rosten@astro.umd.edu}}

\abstract{
Variability is a common characteristic of magnetically active stars.  Flaring
variability is usually interpreted as the observable consequence of transient magnetic
reconnection processes happening in the stellar outer atmosphere.  Stellar flares
have been observed now across 11 decades in wavelength/frequency/energy; such a large span
implies that a range of physical processes takes place during such events.  
Despite the fact that stellar radio flares have long been recognized and studied,
key unanswered questions remain.  I will highlight what, in my opinion, are
some of these questions.
I will also 
describe recent results on stellar flare emissions at radio wavelengths,
discussing the nature of coherent and incoherent emissions and the prospects of
wide-field radio imaging telescopes for studying such events.
          }

\FullConference{Bursts, Pulses and Flickering:Wide-field monitoring of the dynamic radio sky\
		 June 12-15 2007\\
		 Kerastari, Tripolis, Greece}

\begin{document}

\section{Which Stars Flare?}
At first glance, a comparison of some of the titles from the early days
of radio astronomy with some of the targets being talked about
here suggest that not much has changed:  witness %\cite{lovell1963}
Lovell (1963)
``Radio Emission from Flare Stars'' and Bastian et al. (1990)
with the same
title.  Indeed, Sections 3 and 4 of this article
describe radio emission from flare stars (among other classes of stars).
While some of the ``usual suspects'' remain favorite targets for radio
observations due to their interesting nature, the advent of sensitive 
radio interferometers has increased our knowledge of radio emission
and it is now recognized that radio emission can be produced
in the atmospheres of 
many different types of stars.  In particular, as described in G\"{u}del (2002),
 it is now known that stellar radio emission occurs across the
HR diagram, with a variety of manifestations.  Those objects in
the ``cool half'' of the HR diagram, lying roughly near the
main sequence and with spectral types F and later, display predominantly
nonthermal emission.  This nonthermal emission 
is intimately connected with the presence of magnetic fields
whose dynamical re-arrangement 
during magnetic reconnection processes can lead to transient increases
in associated emissions.  
Thus, these cool stellar objects displaying nonthermal emission
are all potential ``flare stars''; the classic moniker must be expanded from the
usual connotation of dMe flare stars to include all stars with evidence for
large-scale surface magnetic fields.
Radio flares of one kind or another have been seen in almost all 
kinds of cool stars displaying other signatures of magnetic activity.\\[-10mm]

\section{Expectations from the Sun}
The Sun is the source of a rich variety of radio emissions which can be
traced to thermal or nonthermal emissions.  Flaring variability is confined
to strong magnetic field regions. 
Even though the Sun is well-studied at radio wavelengths, key 
questions remain.  
One illustration of where further investigation has revealed
new insights into radio flare processes is examining the use of the assumption of
isotropic pitch-angle distribution in gyrosynchrotron emission.
Employing temporal and spectral observations of impulsive solar
radio flares at GHz frequencies, Lee \& Gary (2000)
%\cite{leegary2000} 
have been able to infer some of the properties of 
electrons injected into magnetic traps, finding that an initially narrow
distribution of electron pitch angles is required.
In stellar radio observations, one of the standard assumptions is that
of an isotropic pitch angle distribution; this paper, along with work
on other aspects of gyrosynchrotron emission from solar radio flares
%(Fleishman \& Melnikov 2003)
%\cite{fm2003a,fm2003b}
reveals the Pandora's box 
opened by high quality flare observations, and highlights some 
areas in stellar radio flare studies where the next generation of radio facilities can make progress.

One of the key advances solar radiophysics made in the 1950s
which allowed identification
of the mechanisms involved in metric and decimetric solar radio emissions
was the use of the dynamic spectrum.  Wild et al. (1959),
%\cite{wild1959} 
in particular,
showed that the complex drifting structures present in the frequency-time domain
could be interpreted as the intrinsic drift of beams of electrons
travelling through the upper solar atmosphere, generating plasma radiation
at the ambient electron density.  Such dynamic spectral analyses
have been used to identify the likely location of particle 
acceleration in the solar corona (Aschwanden \& Benz 1997).  
Under the assumption that the observed emission is plasma radiation at
the fundamental or harmonic, produced by a beam of electrons travelling
through the solar atmosphere and generating Langmuir waves at the local electron
density, 
the observed
frequency drift can be related to atmospheric properties and intrinsic motion via:\\
\begin{equation}
\frac{d{\nu}}{dt} = \frac{\partial \nu}{\partial n_{e}} 
\frac{\partial n_{e}}{\partial h} \frac{\partial h}{\partial s}\frac{\partial s}
{\partial t}
\end{equation}
where $\nu$ is the observed frequency at time $t$, $n_{e}$ is the electron density,
$h$ is the radial height in the atmosphere and $s$ is the path length travelled
in the atmosphere. This equation can be simplified under the assumption of
a barometric atmosphere, to $\dot \nu = \nu \cos \theta v_{B}/(2H_{n})$,
where $v_{B}$ is the speed of the exciter whose motion causes the plasma radiation.
Speeds from 0.1--0.5$c$ have been inferred using such analyses (Wild et al. 1959).
%\cite{wild1959}. 
The miriad other kinds of solar radio bursts, particularly at low
frequencies, testifies to the complexity of solar radio emissions.
I have highlighted here two examples, which are illustrative of the kinds of
observations stellar radio astronomers can use to study similar processes
in stellar environments.

Should we expect radio emission from the most active stars to behave like the Sun?
Intense radio emission is found to be correlated over ten orders of magnitude
with X-ray emission in a sample which includes active stars and 
individual solar flares (Benz \& G\"{u}del 1994).  %While a detailed
%explanation for such a correlation has not been proposed, the 
The correlation points out the necessity for strong magnetic fields
to both accelerate electrons and heat plasma, 
and that the basic phenomenology must hold despite physically
different length scales, gravities, and interior conditions 
found in these vastly different stars (T Tauri stars, coronal giant stars,
active binary systems, dMe flare stars)
compared with the Sun.
The Sun's corona is well-described by relatively cool high
temperature plasma, with T$_{\rm cor} \sim$ 2$\times$10$^{6}$K, with 
electron acceleration localized in space and time, appearing in regions
associated with large magnetic field strengths, and only transiently.
In contrast, most active stars have a distribution of coronal plasma
which peaks at hotter temperates, generally 8--10 $\times$10$^{6}$K
(G\"{u}del 2004)
%\cite{gudel2004} 
and extending up to 40$\times$10$^{6}$K in periods of quiescence.
Electron acceleration, as diagnosed by nonthermal radio emission, appears
to be persistent (Chiuderi Drago \& Franciosini 1993) and taking place on a global scale
(Franciosini \& Chiuderi Drago 1995),
%\cite{franciosini1995}, 
in addition to transient acceleration episodes
identified as flares.  Even if the same basic phenomena are at 
work (magnetic heating of coronal loops, etc.) it is clear these
atmospheres will have different effects on free-free and gyroresonant
opacities ($\kappa_{gr}\sim$T$^{s-1}$, $\kappa_{ff}\sim$T$^{-3/2}$),
and that the hotter coronae favor escape of plasma radiation
(White \& Franciosini 1995).\\[-10mm]

\section{Incoherent Radio Flares}
The most common kind of stellar radio flare is characterized by transient 
increases in flux density on timescales of tens of minutes up to days or longer,
and is associated with low degrees of circular polarization.  The flare temporal evolution 
is similar at adjacent frequency bands, pointing to optical depth effects 
in modifying the emergent radiation.  Such flares can be associated with
X-ray flares (see G\"{u}del et al. 1996, Osten et al. 2004)
although the assocation is not 100\% (see discussion in 
Smith et al. 2005, Osten et al. 2004).
%\cite{smith2005,osten2004}).
These incoherent radio flares are generally attributed to gyrosynchrotron emission
from mildly relativistic electrons (Dulk 1985).% \cite{dulk1985}. 
Due to sensitivity constraints, most radio stellar flare 
observations have taken place at GHz frequencies.
Multi-frequency observations have
illuminated some of these key points.  
The emission outside of the flare
can attain a moderate amount of circular polarization.  A simple model in 
which there is a constant level of flux and circular
polarization (corresponding to quiescence) plus a varying amount of flux which is intrinsically unpolarized
can fit the temporal variability of circular polarization during some 
large-scale flares in which an inverse relationship between flux and circular polarization
is noted (Osten et al. 2004 and Figure~1, left).
%\cite{osten2004} and Figure 1, left).  
During the rising phase of the flare
the spectral index increases, becoming large and positive, 
(Osten et al. 2004, 2005
%(\cite{osten2004,osten2005}
and Figure 1); with spectral index defined 
so that $S_{\nu} \propto \nu^{\alpha}$, this can be interpreted as 
optically thick conditions in the flare rise with a decrease back to optically
thin conditions as the flare progresses.

%\begin{figure}[h]
%\includegraphics[scale=0.3]{stellarneupert.ps}
%\caption{put caption here}
%\end{figure}

\begin{figure}[h]
\begin{center}
\includegraphics[scale=0.5]{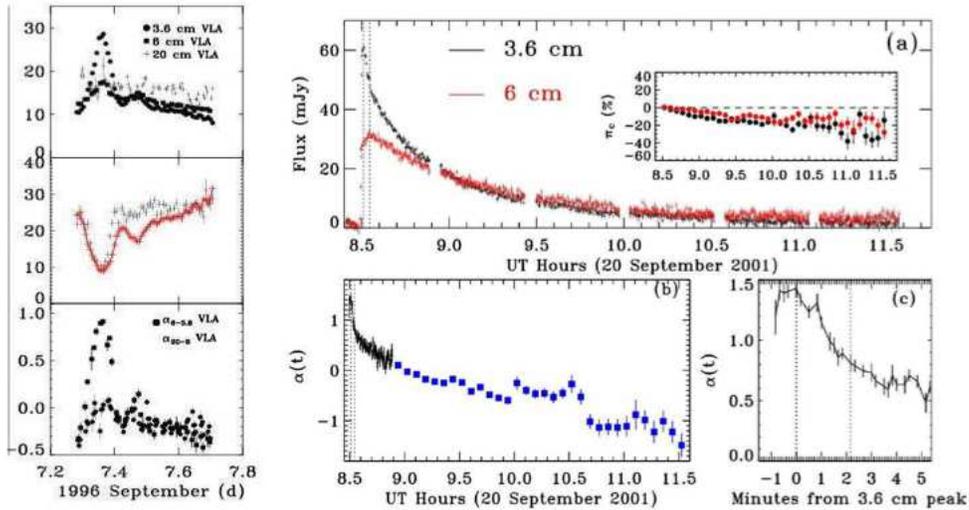}
\caption{{\bf (left)} Multi-frequency radio flare on the active binary system
HR~1099 (V711~Tau) which illustrates the circular polarization and spectral index trends
discussed in the text.  Top panel displays flux variation in mJy at three frequencies; 
middle panel displays temporal trend of observed percent circular polarization
(plusses) with a model for the transient increase in flux density being unpolarized;
bottom panel displays temporal trend of spectral index; the positive correlation with 
flux density suggests optical depth effects occurring during the flare.
Taken from Osten et al. (2004). %\cite{osten2004}.
{\bf (right)} Spectral index and circular polarization variations during a large
flare outburst observed on the dMe flare star EV~Lacertae (Osten et al. 2005).
%\cite{osten2005}.
At flare peak there is no measurable polarization, and the spectral index changes
in the first minutes of the flare point to complex  and fast dynamical changes in 
the flaring plasma.
}
\end{center}
\end{figure}

The largest stellar radio flares at cm wavelengths illustrate some extreme
properties:  peak radio luminosities exceeding 100$\times$ the quiescent 
radio luminosities, lasting for tens of days, and with recurrence timescales
on the order of several times per year (Richards et al. 2003).%\cite{richards2003}.  
Daily monitoring of several active stars, 
which revealed the frequency of these
events, has occurred primarily at cm wavelengths, and the sparse
data collection allows only cursory examination of the flaring process
which can produce these events.  Pointed observations only rarely
pick up these events; 
VLA and VLBA observations from a serendipitously detected superflare
(Beasley \& Bastian 1998) showed that the peak frequency exceed 20 GHz on 5 of the
7 consecutive days during which the flare was observed.
Based on expectations from solar flares, the accelerated electrons producing
the nonthermal radio emission should also produce nonthermal hard X-ray emission,
but until recently there had not been a confident claim of detection of such
emission, despite the reportings of large X-ray flares.
%While radio superflares have been caught in these daily monitoring observations,
%their X-ray counterparts have not been studied due to their rarity and
%the low likelihood of a pointed X-ray observation (lasting typically 100 ks)
%observing one of these.  
%BeppoSAX observations, with instruments covering two orders of magnitude
%in X-ray energy,  did catch several large
%stellar flare enhancements (Favata \& Schmitt 1999, 
%Franciosini et al. 2001),
%\cite{favataschmitt1999,franciosini2001}, 
%but the  X-ray emission was always consistent
%with enhanced thermal emission (up to 10$^{8}$K).  
A serendipitously detected stellar X-ray superflare from the nearby active
binary II~Peg with the Swift satellite (Osten et al. 2007)
%\cite{osten2007} 
displayed the first 
evidence for nonthermal hard X-ray emission, with detections out
to 200 keV.  This allowed the electron density
distribution to be investigated in a stellar flare for the first time.
An estimation of the flare energetics revealed rough agreement between 
nonthermal and thermal energy estimates, with
evidence for continued heating (and particle acceleration) in the decay phase of the flare.
This was only one flare, and it lacked multi-wavelength coordination; future attempts
to obtain X-ray and radio coverage of such events are the stellar astronomers'
best hope for making significant advances in the area of stellar flare particle acceleration.

{\bf Key unanswered questions in incoherent stellar radio flares:\\}
%\begin{enumerate}
%Solar flare studies reveal a rough equipartition between 
%nonthermal flare energy estimates and thermal flare energ estimates.
%Nonthermal energy estimates must be made using indicators free of 
%interpretational bias.
%\item 
\noindent {\bf 1. Can detailed spectral and temporal observations of stellar radio
flares across a large frequency range be used to determine the evolution of particle
acceleration and investigate particle injection?  }
Observations which are clearly in the optically thin regime are needed
to deconvolve spectral index evolution into optical depth evolution
(which involves changing magnetic fields and particle distribution) and
evolution of the particle distribution.\\
%\item 
\noindent {\bf 2. Is there evidence for particle trapping or pitch-angle dependence of emission
from stellar radio flares?}
By having detailed spectral and temporal observations, particularly in the optically
thin regime, such effects can be investigated. \\
%\item 
\noindent {\bf 3. How does the nonthermal energy release compare to the rest of the flare
energy budget? }
Answering this question allows investigations into the flare process itself,
the partition of flare energetics, and the timing of particle acceleration
and plasma heating.\\
%\item 
\noindent {\bf 4. How do these things compare to relatively well-studied solar flares, and
how do they vary in different stellar environments?}
Does a flare on a diskless T Tauri star behave the same way as that on a
dMe flare star, or a coronal giant star?  Observations can be used to infer
the physical conditions present in the atmospheres of flaring stars, which can be
important for other reasons (e.g. physical conditions around young stars).
%\end{enumerate}

In order to answer these questions,  a combination of broad-band
coverage and sensitivity is needed, particularly at high frequency.  In addition,
in order to study the stellar ``superflares'', the ability to respond
quickly to external triggers (e.g. from X-ray telescopes) is needed.\\[-10mm]

\section{Coherent Radio Flares}
A key point about stellar radio flares is that they come in
(at least) two flavors:  the largely unpolarized flares described
above, and then another category, 
described by high degrees of circular polarization
(reaching 100\%, something not found in other astrophysical sources of radiation),
with timescales varying from milliseconds up to hours, emission structured
in both frequency and time and no clear association with X-ray flares.
These properties constrain the intrinsic brightness temperatures to 
be in excess of the 10$^{12}$K limit for
incoherent radiation set by inverse Compton scattering 
%(Kellerman \& Pauliny-Toth 1969) %\cite{kellerman1969} 
and thus must be coherent in nature.
Study of other classes of active stars have revealed evidence for
similar kinds of transient coherent emission.  The frequencies at which
such emission has been detected range from metric wavelengths 
(Kundu \& Shevgaonkar 1988) %\cite{kundu1988}
to centimeter wavelengths (Bingham et al. 2001). %\cite{bingham2001}.
Coherent bursts appear structured in time and frequency.
This was noted in previous observations of flares from the classic flare
stars (Bastian et al. 1990),
%\cite{bastianetal1990}, 
although the ability to study such events in detail had been limited by a combination
of bandwidth coverage and time resolution. 
%\cite{gudel1989} 
G\"{u}del et al. (1989) used
three single-dish telescopes to determine that the radio bursts often observed
on AD~Leo were in fact extraterrestrial and covered a large frequency range,
but quantitative analysis of burst properties was elusive.
On the Sun there are two distinct coherent emission mechanisms at work:
one being associated with plasma radiation (operating at the fundamental
or harmonics of the plasma frequency) and the other associated with
an electron cyclotron maser (operating at the fundamental or harmonic of
the electron cyclotron frequency).  The quandary with radio bursts seen on 
flare stars lay in not being able to prefer one mechanism to the other,
based on the available observational discriminants, usually
consisting of lower limits to brightness temperature, and observed degrees
of circular polarization.

The advent of wide bandwidth radio instruments has revolutionized the
arena of coherent stellar radio bursts.  By combining 
large frequency bandwidth ratios with high time resolution, 
instrumental backends originally designed for use in studying pulsars
can be harnessed to a quantitative study of stellar radio burst characteristics.
In a study by Osten \& Bastian (2006),
%\cite{ob2006}, 
timescales, 
instantaneous bandwidths, and frequency drifts of a number of bursts could be measured,
along with flux densities and circular polarizations.  The observed
properties of the bursts, in concert with ancillary knowledge about the
stellar atmospheric properties, led the authors to prefer the plasma radiation
mechanism. 

\begin{figure}[h]
\begin{center}
\includegraphics[scale=0.6]{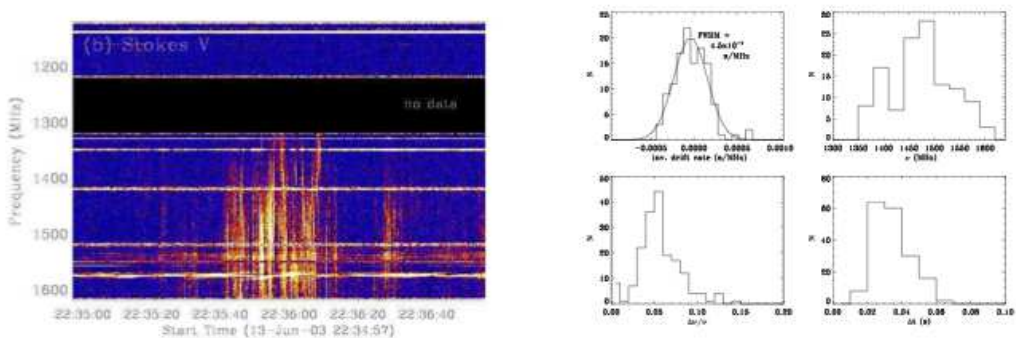}
\caption{{\bf (left)} Dynamic spectrum of stellar radio burst observed with Arecibo
on the flare star AD~Leo (Osten \& Bastian 2006).
%\cite{ob2006}.  
The burst illustrates several
key points about stellar coherent flare emission:  large amounts of circular polarization
(here $>$ 90\%); evidence for structure in the frequency and time domains.  The
time resolution here of 10 ms allowed lower limits on brightness temperature of 10$^{14}$K.
{\bf (right)} Histograms of burst properties deduced from analysis of the dynamic spectrum.  
Important burst properties that could be constrained using such observations include
the drift rate of the bursts, instantaneous bandwidth ratio, time duration, and start frequency.
Based on these properties, Osten \& Bastian concluded that the bursts were likely produced
by plasma radiation, and were the stellar equivalent of decimetric solar type III bursts.
}
\end{center}
\end{figure}

Further observations with even higher time resolution (Osten \& Bastian 2007) %\cite{ob2007}
have revealed
additional behaviors.  The appearance of drifting structures in the
dynamic spectrum is apparently common, but complex diffuse structures
have also been seen.
These highest time resolution observations of stellar radio bursts
indicate burst timescales of
several ms, indicating a lower limit to the brightness temperature
of 10$^{18}$K.  This places severe constraints on the emission process
in this instance; based on the observed properties of the drifting structures
seen here in comparison with those seen in earlier observations, a cyclotron
maser emission may be preferred.  
Osten \& Bastian ruled out other causes of the
dynamic spectral structure, such as that induced by stellar
atmospheric properties, or scattering in the stellar corona or intervening
interstellar medium.

Currently only
the largest flares are amenable to such high time-resolution, wide bandwidth
dynamic spectral analysis, but coherent flares are detected across a range of
radio luminosities and from almost all classes of magnetically active stars.
{\em Coherent emission is a manifestation
of magnetic activity which has hitherto been unexploited for what it can reveal
about stellar atmospheric properties.}
New generations of radio telescopes will probe the low frequency regime, where
coherent emissions are expected to be richer and more varied.
Although I have concentrated primarily on flare stars in this brief discussion,
recent work in other areas has shown the importance of coherent processes in
producing stellar radio emission -- the pulstar of Kellett et al. (2007),%\cite{kellett2007},
and the highly circularly polarized rotationally modulated emission from some
very low mass stars and brown dwarfs (Hallinan et al. 2007).
%\cite{hallinan2006,hallinan2007}.
%The most common categories of active stars which have been reported to
%show coherent emission are the classic dMe flare stars and active
%binary systems, but other classes of active stars show hints of 
%producing this emission as well (T Tauri stars [ref], coronal giants [ref],
%very low mass dwarfs [ref]).  

{\bf Key unanswered questions in coherent stellar flares:}\\%[-8mm]
%\begin{enumerate}
\noindent
%\item 
{\bf 1. What is the nature of coherent emission in different stellar systems?}
The observational manifestation is large amounts of circular polarization which are
transient on some timescale. As has been noted in previous papers on this subject
(Bastian et al. 1990),
%\cite{bastianetal1990},
there is an ambiguity as to whether such emissions could be produced by a plasma
emission process or a cyclotron maser process.  Based on consideration of both
the stellar atmospheric properties, often largely unknown, and more detailed
observational constraints on the coherent stellar radio bursts, it may be possible
to determine whether particular types of coherent processes preferentially
operate in particular
stellar environments. \\
\noindent 
%\item 
{\bf 2. What is the relation of coherent flaring to flaring in other wavelength
regions?}
Coherent emission generally involves energetic particles, 
and solar studies have shown correlations between nonthermal X-ray flare signatures 
and electron beam-excited radio emissions (Aschwanden \& Benz 1997).
Previous studies
on stellar flares have tended to find no clear association between coherent flares
and flare signatures at other wavelengths (Kundu et al. 1988).  
Determining whether an association exists will be enormously helpful in 
clarifying some of the processes involved in stellar flares.\\
\noindent 
%\item 
{\bf 3. What do 
analyses of flare structure in the frequency and time domains  
imply about constraints on stellar atmospheric (particularly B, $n_{e}$) structuring?}
The ability of dynamic spectral analyses to reveal crucial information about
the densities and magnetic field strengths at various points in the
atmosphere in which radio emission from sub-bursts is produced allows a probe of the physical
conditions in stellar atmospheres which cannot be done using techniques at other
wavelength regions, where often the disk-averaged emission and density-squared dependence
of the thermal emission
introduce biases.  This can be especially important in areas such as 
using coherent radio emission from T Tauri stars to probe the star-disk connection.\\[-10mm]
%\end{enumerate}

\section{Implications for Wide-Field Imaging}
It is curious, given the large flux densities recorded in 
large stellar outburst from pointed radio observations,
that unbiased radio surveys apparently don't find many radio stars.
As an example, the FIRST survey (Helfand et al. 1999)
%\cite{helfand1999} 
surveyed 5000 square degrees down to 
a 0.7 mJy limit at 20 cm and found 26 stellar radio sources, with only
16 of those being new detections.  Some of the limiting factors to survey
sensitivity for transient emission include
area coverage, sensitivity, and time on field, all of which increase the likelihood
of catching transient behavior.  
Based on the known properties of nearby 
radio-active stars, we can inform some expectations for wide-field imaging that
the next generation of radio telescopes (e.g. ATA, LOFAR, MWA, LWA)
can expect to see.  I concentrate here on M dwarf flare stars and active binary systems
due to the fact that more is known about their behavior. 

M dwarf flare stars can have large coherent bursts, up to 500 times the
quiescent flux density levels, reaching Jy-level flux densities for
the nearest studied objects. 
These events have been observed at decimeter and meter wavelengths.
Telescopes with sub-mJy sensitivity will be able to catch
similar events from dMe flare stars at distances up to $\approx$ 150 pc.
These 
high intensity events are relatively rare, however, with duty cycles
$<$ 1\%.
Incoherent bursts tend to be smaller in amplitude.
40\%  of surveyed flare stars were detected at 20 and 3.6 cm (White et al. 1989).
%\cite{white1989}.
If they all behave the same way this is a large reservoir of stellar transient
radio sources which, given the enhancements, can be seen out to large galactic
distances.
An order of magnitude estimate of the number of flare stars visible with
telescopes of sub-mJy sensitivity can be made by computing $n 4\pi d^{3}/3 f_{R}$,
where $n$ is the space density of flare stars (0.08 stars per pc$^{3}$,
Reid et al. 2007),
%\cite{reid2007}),
$d$ is the maximum distance to which telescopes are sensitive using the general
numbers above, and $f_{R}$ the
fraction of radio-emitting objects. Given the parameters
above, 4.5$\times$10$^{5}$ sources are expected, a potential
foreground ``fog'' of flare stars (Kulkarni \& Rau 2006). %\cite{dmefog}. 
If these objects are emitting high brightness temperature bursts
$<$1\% of the time, as many as 4500 flare stars may be observable.
Gyrosynchrotron flares from dMe flare stars usually have smaller enhancements,
(the largest being factors of tens) so these will be visible out to smaller distances, but
may happen more often relatively speaking.  Currently neither
the distribution of occurrence of coherent bursts at a particular frequency,
nor the distribution of incoherent flares, is constrained, lending 
considerable uncertainties to these expectations.
Better constraints on the expected behavior of coherent bursts from dMe
flare stars, as well as good positional accuracy of transient radio
sources, will be needed to make identifications with flare stars.

Based on the large amplitude incoherent flares, as well as the
relatively large amplitude coherent bursts observed on active binary
systems, these objects are potentially visible at larger distances than
dMe flare stars.
Coherent emission from active binary systems appears to be common: from 
several epochs observed at 1.4 GHz on HR 1099, moderately or highly circularly polarized
emission was observed $\sim$33\% of the time (Osten et al. 2004). 
%\cite{osten2004}.
%X-ray flares are also common, occuring $\sim$1/3 of the time, and so 
%associated incoherent flares may be expected to be similarly common.
Enhancements in gyrosynchrotron flares of more than
100 times the quiescent levels at cm wavelengths and shorter have been observed,
with recurrence times of several per year.
X-ray observations of active binaries reveal that they undergo large
X-ray flares about 1/3 of the time they are observed; similar statistics are
not available for gyrosynchrotron flares. 
There are 206 active binaries catalogued within $\sim$200 pc
(Strassmeier et al. 1993); %\cite{cabs2} 
Drake et al. (1989) %\cite{drake1989}
detected 66/122 at 5 GHz.
Using the same metric as above, for $n$=4--8$\times$10$^{-5}$
active binaries per pc$^{3}$ (Favata et al. 1995),
%\cite{favata1995}, 
$f_{R}$=0.54, Jy-level flares detected
with sub-mJy sensitivity will 
be detectable at distances up to $\sim$150 pc, leading to $\approx$300--600
sources.  Similar conditions apply here as in transient dMe flare star 
radio emission to identify transient radio sources as an active binary system.\\[-8mm]

\section{Conclusions}
Targeted observations of nearby active stars suggest good chances for 
stellar detections with the next generation wide-field radio surveys.
Positional accuracy is needed; with variability, the timescale, inferred
luminosities, and parameters of
the transient radio emission can help identify stellar transients.  
Targeted observations are
still the best way to make progress in understanding the details of
flare dynamics.  Large area surveys have the best potential to reveal
large enhancement and/or serendipitous stellar flares due to long effective on-source times,
and reveal whether the behavior seen in targeted observations is typical
of the larger source population.

\newcommand{\apj}{Astrophysical Journal }
\newcommand{\aap}{Astronomy \& Astrophysics }
\newcommand{\aaps}{Astronomy \& Astrophysics Supplement}
\newcommand{\nat}{Nature }
\newcommand{\solphys}{Solar Physics }
\newcommand{\araa}{Annual Reviews of Astronomy and Astrophysics}
\newcommand{\apjs}{Astrophysical Journal Supplement}
\newcommand{\aapr}{Astronomy and Astrophysics Review}
\newcommand{\apjl}{Astrophysical Journal Letters}
\newcommand{\aj}{Astronomical Journal}

%\bibliographystyle{natbib}
%\bibliography{/export/users/rosten/procs/pos}
%\include{skeleton.bbl}
%\begin{thebibliography}{42}
%\expandafter\ifx\csname natexlab\endcsname\relax\def\natexlab#1{#1}\fi

\noindent {\bf References}
 
%\bibitem[{{Aschwanden} \& {Benz}(1997)}]{ab1997}
{Aschwanden}, M.~J. \& {Benz}, A.~O. \emph{Electron Densities in Solar Flare Loops, Chromospheric Evaporation Upflows, and Acceleration Sites} 1997, \apj, 480, 825
 
%\bibitem[{{Bastian}(1990)}]{bastian1990}
{Bastian}, T.~S. \emph{Radio emission from flare stars} 1990, \solphys, 130, 265
 
%\bibitem[{{Bastian} {et~al.}(1990){Bastian}, {Bookbinder}, {Dulk}, \&
  %{Davis}}]{bastianetal1990}
%{Bastian}, T.~S., {Bookbinder}, J., {Dulk}, G.~A., \& {Davis}, M. 1990, \apj,
{Bastian}, T.~S. et al. \emph{Dynamic spectra of radio bursts from flare stars} 1990, \apj,
  353, 265
 
%\bibitem[{{Beasley} \& {Bastian}(1998)}]{bb1998}
{Beasley}, A.~J. \& {Bastian}, T.~S. \emph{VLBA Imaging of UX Ari} 1998, in {\emph Astronomical Society of the
  Pacific Conference Series, Vol. 144, IAU Colloq. 164: Radio Emission from
  Galactic and Extragalactic Compact Sources}, ed. J.~A. {Zensus}, G.~B.
  {Taylor}, \& J.~M. {Wrobel}, 321
 
%\bibitem[{{Benz} \& {G\"{u}del}(1994)}]{bg1994}
{Benz}, A.~O. \& {G\"{u}del}, M. \emph{X-ray/microwave ratio of flares and coronae
} 1994, \aap, 285, 621
                                                                                                    
%\bibitem[{{Bingham} {et~al.}(2001){Bingham}, {Cairns}, \&
  %{Kellett}}]{bingham2001}
{Bingham}, R., {Cairns}, R.~A., \& {Kellett}, B.~J. \emph{Coherent cyclotron maser radiation from UV Ceti} 2001, \aap, 370, 1000
                                                                                                    
%\bibitem[{{Chiuderi Drago} \& {Franciosini}(1993)}]{chiuderi1993}
{Chiuderi Drago}, F. \& {Franciosini}, E. \emph{Flaring and quiescent radio emission of UX ARIETIS - A time-dependent model} 1993, \apj, 410, 301
                                                                                                    
%\bibitem[{{Drake} {et~al.}(1989){Drake}, {Simon}, \& {Linsky}}]{drake1989}
{Drake}, S.~A., {Simon}, T., \& {Linsky}, J.~L. \emph{A survey of the radio continuum emission of RS Canum Venaticorum and related active binary systems} 1989, \apjs, 71, 905
                                                                                                    
%\bibitem[{{Dulk}(1985)}]{dulk1985}
{Dulk}, G.~A. \emph{Radio emission from the sun and stars} 1985, \araa, 23, 169
                                                                                                    
%\bibitem[{{Favata} {et~al.}(1995){Favata}, {Micela}, \&
  %{Sciortino}}]{favata1995}
{Favata}, F., {Micela}, G., \& {Sciortino}, S. \emph{The space density of active binaries from X-ray surveys.} 1995, \aap, 298, 482
                                                                                                    
%\bibitem[{{Favata} \& {Schmitt}(1999)}]{favataschmitt1999}
%{Favata}, F. \& {Schmitt}, J.~H.~M.~M. \emph{Spectroscopic analysis of a super-hot giant flare observed on Algol by BeppoSAX on 30 August 1997} 1999, \aap, 350, 900

%\bibitem[{{Fleishman} \& {Melnikov}(2003{\natexlab{b}})}]{fm2003a}
%{Fleishman}, G.~D. \& {Melnikov}, V.~F. \emph{Optically Thick Gyrosynchrotron Emission from Anisotropic Electron Distributions} 2003a, \apj, 584, 1071

%\bibitem[{{Fleishman} \& {Melnikov}(2003{\natexlab{a}})}]{fm2003b}
%{Fleishman}, G.~D. \& {Melnikov}, V.~F. \emph{Gyrosynchrotron Emission from Anisotropic Electron Distributions} 2003b, \apj, 587, 823
                                                                                                    
%\bibitem[{{Franciosini} \& {Chiuderi Drago}(1995)}]{franciosini1995}
{Franciosini}, E. \& {Chiuderi Drago}, F. \emph{Radio and X-ray emission in stellar magnetic loops.}1995, \aap, 297, 535
                                                                                                    
%\bibitem[{{Franciosini} {et~al.}(2001){Franciosini}, {Pallavicini}, \&
  %{Tagliaferri}}]{franciosini2001}
{Franciosini}, E., {Pallavicini}, R., \& {Tagliaferri}, G. \emph{BeppoSAX observation of a large long-duration X-ray flare from UX Arietis} 2001, \aap, 375, 196
                                                                                                    
%\bibitem[{{G{\"u}del}(2002)}]{gudel2002}
{G{\"u}del}, M. \emph{Stellar Radio Astronomy: Probing Stellar Atmospheres from Protostars to Giants} 2002, \araa, 40, 217
                                                                                                    
%\bibitem[{{G{\"u}del}(2004)}]{gudel2004}
---. \emph{X-ray astronomy of stellar coronae} 2004, \aapr, 12, 71

%\bibitem[{{G\"{u}del} {et~al.}(1989){G\"{u}del}, {Benz}, {Bastian}, {Furst},
  %{Simnett}, \& {Davis}}]{gudel1989}
{G\"{u}del}, M. et al. \emph{Broadband spectral observation of a dMe star radio flare} 
  1989, \aap, 220, L5
                                                                                                    
%\bibitem[{{G\"{u}del} {et~al.}(1996){G\"{u}del}, {Benz}, {Schmitt}, \&
%  {Skinner}}]{gudel1996}
{G\"{u}del}, M., et al. \emph{The Neupert Effect in Active Stellar Coronae: Chromospheric Evaporation and Coronal Heating in the dMe Flare Star Binary UV Ceti} 
  1996, \apj, 471, 1002
                                                                                                    
%\bibitem[{{Hallinan} {et~al.}(2006){Hallinan}, {Antonova}, {Doyle}, {Bourke},
%  {Brisken}, \& {Golden}}]{hallinan2006}
%{Hallinan}, G., et al. \emph{Rotational Modulation of the Radio Emission from the M9 Dwarf TVLM 513-46546: Broadband Coherent Emission at the Substellar Boundary?}
%  A. 2006, \apj, 653, 690
                                                                                                    
%\bibitem[{{Hallinan} {et~al.}(2007){Hallinan}, {Bourke}, {Lane}, {Antonova},
%  {Zavala}, {Brisken}, {Boyle}, {Vrba}, {Doyle}, \& {Golden}}]{hallinan2007}
{Hallinan}, G., et al. \emph{Periodic Bursts of Coherent Radio Emission from an Ultracool Dwarf}
  2007, \apjl, 663, L25

%\bibitem[{{Helfand} {et~al.}(1999){Helfand}, {Schnee}, {Becker}, {White}, \&
%  {McMahon}}]{helfand1999}
{Helfand}, D.~J. et al. \emph{The FIRST Unbiased Survey for Radio Stars}
 1999, \aj, 117, 1568
                                                                                                    
%\bibitem[{{Kellermann} \& {Pauliny-Toth}(1969)}]{kellerman1969}
{Kellermann}, K.~I. \& {Pauliny-Toth}, I.~I.~K. \emph{The Spectra of Opaque Radio Sources} 1969, \apjl, 155, L71+
                                                                                                    
%\bibitem[{{Kellett} {et~al.}(2007){Kellett}, {Graffagnino}, {Bingham},
%  {Muxlow}, \& {Gunn}}]{kellett2007}
{Kellett}, B.~J. et al. \emph{CU Virginis - The First Stellar Pulsar}
2007, {\tt astro-ph/0701214} 
                                                                                                    
%\bibitem[{{Kulkarni} \& {Rau}(2006)}]{dmefog}
{Kulkarni}, S.~R. \& {Rau}, A. \emph{The Nature of the Deep Lens Survey Fast Transients
} 2006, \apjl, 644, L63
                                                                                                    
%\bibitem[{{Kundu} \& {Shevgaonkar}(1988)}]{kundu1988}
{Kundu}, M.~R. \& {Shevgaonkar}, R.~K. \emph{Detection of the dMe flare star YZ Canis Minoris simultaneously at 20 and 90 centimeter wavelengths} 1988, \apj, 334, 1001
                                                                                                    
%\bibitem[{{Kundu} {et~al.}(1988){Kundu}, {White}, {Jackson}, \&
%  {Pallavicini}}]{kunduetal1988}
{Kundu}, M.~R., {White}, S.~M., {Jackson}, P.~D., \& {Pallavicini}, R. 
  \emph{Co-ordinated VLA and EXOSAT observations of the flare stars UV Ceti, EQ Pegasi, YZ Canis Minoris, and AD Leonis} 1988,
  \aap, 195, 159
                                                                                                    
%\bibitem[{{Lee} \& {Gary}(2000)}]{leegary2000}
{Lee}, J. \& {Gary}, D.~E. \emph{Solar Microwave Bursts and Injection Pitch-Angle Distribution of Flare Electrons} 2000, \apj, 543, 457
                                                                                                    
%\bibitem[{{Lovell}(1963)}]{lovell1963}
{Lovell}, B. \emph{Radio Emission from Flare Stars} 1963, \nat, 198, 228
                                                                                                    
%\bibitem[{{Osten} \& {Bastian}(2006)}]{ob2006}
{Osten}, R.~A. \& {Bastian}, T.~S. \emph{Wide-Band Spectroscopy of Two Radio Bursts on AD Leonis} 2006, \apj, 637, 1016
                                                                                                    
%\bibitem[{{Osten} \& {Bastian}(2007)}]{ob2007}
--- {\em Ultra high time Resolution Observations of Radio Bursts on AD~Leonis} 2007, \apj, accepted
                                                                                                    
%\bibitem[{{Osten} {et~al.}(2004){Osten}, {Brown}, {Ayres}, {Drake},
%  {Franciosini}, {Pallavicini}, {Tagliaferri}, {Stewart}, {Skinner}, \&
%  {Linsky}}]{osten2004}
{Osten}, R.~A. et al. \emph{A Multiwavelength Perspective of Flares on HR 1099: 4 Years of Coordinated Campaigns}
  2004, \apjs, 153, 317
                                                                                                    
%\bibitem[]{osten2005}
{Osten}, R.~A., et al.\emph{From Radio to X-Ray: Flares on the dMe Flare Star EV Lacertae}
  2005, \apj, 621, 398

%\bibitem[{{Osten} {et~al.}(2007){Osten}, {Drake}, {Tueller}, {Cummings},
  %{Perri}, {Moretti}, \& {Covino}}]{osten2007}
{Osten}, R.~A., et al. \emph{Nonthermal Hard X-ray Emission and Iron K$\alpha$ Emission
from a Superflare on II~Pegasi}
  2007, \apj, 654, 1052
                                                                                                    
%\bibitem[]{reid2007}
{Reid}, I.~N., {Cruz}, K.~L., \& {Allen}, P.~R. \emph{Meeting the Cool Neighbors. XI. Beyond the NLTT Catalog} 2007, \aj, 133, 2825
                                                                                                    
%\bibitem[]{richards2003}
{Richards}, M.~T., \emph{Statistical Analysis of 5 Year Continuous Radio Flare Data from
$\beta$ Persei, V711 Tauri, $\delta$ Librae and UX Arietis}
  2003, \apjs, 147, 337
                                                                                                    
%\bibitem[]{smith2005}
{Smith}, K., {G{\"u}del}, M., \& {Audard}, M. \emph{Flares observed with XMM-Newton and the VLA
} 2005, \aap, 436, 241
                                                                                                    
%\bibitem[]{cabs2}
{Strassmeier}, K.~G., et al. \emph{A catalog of chromospherically active binary stars (second edition)} 1993,
  \aaps, 100, 173
                                                                                                    
%\bibitem[]{white1995}
{White}, S.~M. \& {Franciosini}, E. \emph{Circular polarization in the radio emission of RS canum venaticorum binaries} 1995, \apj, 444, 342
                                                                                                    
%\bibitem[]{white1989}
{White}, S.~M., {Jackson}, P.~D., \& {Kundu}, M.~R. \emph{A VLA survey of nearby flare stars} 1989, \apjs, 71, 895
                                                                                                    
%\bibitem[]{wild1959}
{Wild}, J.~P., {Sheridan}, K.~V., \& {Neylan}, A.~A. \emph{An Investigation of the Speed of the Solar Disturbances responsible for Type III Radio Bursts} 1959, Australian Journal
  of Physics, 12, 369
                                                                                                    
%\end{thebibliography}

\end{document}